\documentclass{article}
\usepackage{spconf,amssymb,amsmath,graphicx,booktabs}
\usepackage{multirow}
\usepackage{makecell}
\usepackage{url}
\usepackage{hyperref}

\PassOptionsToPackage{hyphens}{url}\usepackage{hyperref}

\usepackage{xcolor}
\usepackage{paralist}
\usepackage{enumitem}
\usepackage[caption=false]{subfig}
\usepackage{float}

\usepackage{multicol}
\usepackage{tabularx}
\usepackage{color}
\usepackage{booktabs}
\usepackage[dvips]{epsfig}
\usepackage{setspace}
\usepackage[utf8]{inputenc} 
\usepackage{appendix}

\usepackage[
backend=biber,
style=ieee,
maxbibnames=3,
maxcitenames=3,
doi=true,isbn=false,url=false,eprint=false
]{biblatex}
\defbibheading{bibliography}[\refname]{}

\DeclareSourcemap{
	\maps[datatype=bibtex, overwrite=true]{
		\map{
			\step[fieldsource=booktitle,
			match=\regexp{.*Interspeech.*},
			replace={Proc. Interspeech}]
			\step[fieldsource=journal,
			match=\regexp{.*INTERSPEECH.*},
			replace={Proc. Interspeech}]
			\step[fieldsource=booktitle,
			match=\regexp{.*ICASSP.*},
			replace={Proc. ICASSP}]
			\step[fieldsource=booktitle,
			match=\regexp{.*icassp_inpress.*},
			replace={Proc. ICASSP (in press)}]
			\step[fieldsource=booktitle,
			match=\regexp{.*International.*Conference.*on.*Acoustics,.*Speech.*and.*Signal.*Processing.*},
			replace={Proc. ICASSP}]
			\step[fieldsource=booktitle,
			match=\regexp{.*International.*Conference.*on.*Learning.*Representations.*},
			replace={Proc. ICLR}]
			\step[fieldsource=booktitle,
			match=\regexp{.*International.*Conference.*on.*Machine.*Learning.*},
			replace={Proc. ICML}]
			\step[fieldsource=booktitle,
			match=\regexp{.*International.*conference.*on.*machine.*learning.*},
			replace={Proc. ICML}]
			\step[fieldsource=booktitle,
			match=\regexp{.*Automatic.*Speech.*Recognition.*and.*Understanding.*},
			replace={Proc. ASRU}]
			\step[fieldsource=booktitle,
			match=\regexp{.*Spoken.*Language.*Technology.*},
			replace={Proc. SLT}]
			\step[fieldsource=booktitle,
			match=\regexp{.*Speech.*Synthesis.*Workshop.*},
			replace={Proc. SSW}]
			\step[fieldsource=booktitle,
			match=\regexp{.*workshop.*on.*speech.*synthesis.*},
			replace={Proc. SSW}]
			\step[fieldsource=booktitle,
			match=\regexp{.*Advances.*in.*neural.*information.*processing.*},
			replace={Proc. NeurIPS}]
			\step[fieldsource=booktitle,
			match=\regexp{.*Advances.*in.*Neural.*Information.*Processing.*},
			replace={Proc. NeurIPS}]
			\step[fieldsource=booktitle,
			match=\regexp{.*Workshop.*on.* Applications.* of.* Signal.*Processing.*to.*Audio.*and.*Acoustics.*},
			replace={Proc. WASPAA}]
			\step[fieldsource=booktitle,
			match=\regexp{.*ISMIR.*},
			replace={Proc. ISMIR}]   
			\step[fieldsource=publisher,
			match=\regexp{.+},
			replace={{}}]
			\step[fieldsource=month,
			match=\regexp{.+},
			replace={{}}]
			\step[fieldsource=location,
			match=\regexp{.+},
			replace={{}}]
			\step[fieldsource=address,
			match=\regexp{.+},
			replace={{}}]
			\step[fieldsource=organization,
			match=\regexp{.+},
			replace={{}}]
		}
	}
}

\newcommand{\redit}[1]{\textcolor{black}{#1}}

\newcommand{\rredit}[1]{\textcolor{black}{#1}}

\newcommand{\nushort}{T13 }

\title{A Comparative Study of Voice Conversion Models with Large-Scale Speech and Singing Data: The T13 Systems for\\the Singing Voice Conversion Challenge 2023}
\name{Ryuichi Yamamoto$^{1,2}$, Reo Yoneyama$^1$, Lester Phillip Violeta$^1$, Wen-Chin Huang$^1$, Tomoki Toda$^1$}
\address{$^{1}$Nagoya University, Japan, $^{2}$LINE Corp., Japan.}

\copyrightnotice{979-8-3503-0689-7/23/\$31.00~\copyright2023 IEEE}
\begin{document}
\ninept
%
\maketitle
\begin{abstract}
This paper presents our systems (denoted as T13) for the singing voice conversion challenge (SVCC) 2023. For both in-domain and cross-domain \rredit{English} singing voice conversion (SVC) tasks (Task 1 and Task 2), we adopt a recognition-synthesis approach with self-supervised learning-based representation. To achieve data-efficient SVC with a limited amount of target singer/speaker's data (150 to 160 utterances for SVCC 2023), we first train a diffusion-based any-to-any voice conversion model using publicly available large-scale 750 hours of speech and singing data. Then, we finetune the model for each target singer/speaker of Task 1 and Task 2. Large-scale listening tests conducted by SVCC 2023 show that our T13 system achieves competitive naturalness and speaker similarity for the harder cross-domain SVC (Task 2), which implies the generalization ability of our proposed method. Our objective evaluation results show that using large datasets is particularly beneficial for cross-domain SVC.
\end{abstract}
\begin{keywords}
Singing voice conversion challenge, singing voice conversion, voice conversion, self-supervised learning
\end{keywords}
\vspace{-2mm}
\section{Introduction}
\label{sec:intro}
\vspace{-1mm}


Singing voice conversion (SVC) is the task of converting speaker identity of source singing to that of target singing while maintaining linguistic contents unchanged, and considered as a specific application of voice conversion (VC) techniques.
With the rising interests in SVC for entertainment industry, there have been many studies on SVC~\cite{chen2019singing,polyak20b_interspeech,liu2021fastsvc,wang22u_interspeech,jayashankar2023self}.


Owing to the recent advances of deep learning, the current \redit{state-of-the-art} VC systems can generate synthetic speech samples nearly close to the human voice~\cite{lorenzo2018voice,zhao2020voice}. However, there are still challenges in SVC compared to well-studied speech VC: (1) high-quality singing voice dataset is much more difficult to collect than speech.
Even though several works attempted to construct singing databases for research purposes~\cite{wang2022opencpop,zhang2022m4singer,ogawa2021tohoku,tamaru2020jvs,duan2013nus,koguchi2020pjs,sharma2021nhss}, the amount of publicly available singing datasets remains much smaller than that of large-scale speech datasets (e.g., 50 hours for OpenSinger~\cite{huang2021multi}~vs.~1,000 hours for LibriSpeech~\cite{panayotov2015librispeech}).
(2) Furthermore, prosody-related factors such as pitch patterns and timing deviations\redit{, which are part of the singing style,} need to be more carefully converted to preserve the \rredit{underlying} musical score.



In this study, we address the first challenge by investigating SVC models using large-scale speech and singing datasets.
Given that the dataset provided by the singing voice conversion challenge (SVCC) 2023 contains only 150 to 160 short English audio clips for each target singer/speaker~\cite{huang2023singing}, we overcome this limitation by utilizing a diverse blend of publicly available speech and singing datasets \redit{not limited to the English language}.
This approach allows our VC models to generalize to various speakers and singers well.
\redit{
To investigate the effectiveness of the proposed method, we conduct a comparative study with various training data configurations, such as singing only and a mixture of speech/singing, as well as different model types, including ContentVec~\cite{qian2022contentvec} and HuBERT-soft~\cite{van2022comparison}.
Large-scale subjective evaluations conducted by SVCC 2023 show that our best system (denoted as T13) achieves competitive naturalness and speaker similarity for the harder cross-domain SVC (Task~2), which implies the generalization ability of our method.
Our objective evaluation results confirm that using a large amount of datasets is particularly beneficial for cross-domain SVC. 
Audio samples are available on our demo page~\footnote{
\label{demo}\url{https://r9y9.github.io/projects/svcc2023/}
}.}

\vspace{-2mm}
\section{Summary of our T13 systems for SVCC 2023}
\vspace{-1mm}

\begin{figure*}[t]
  \centerline{\epsfig{figure=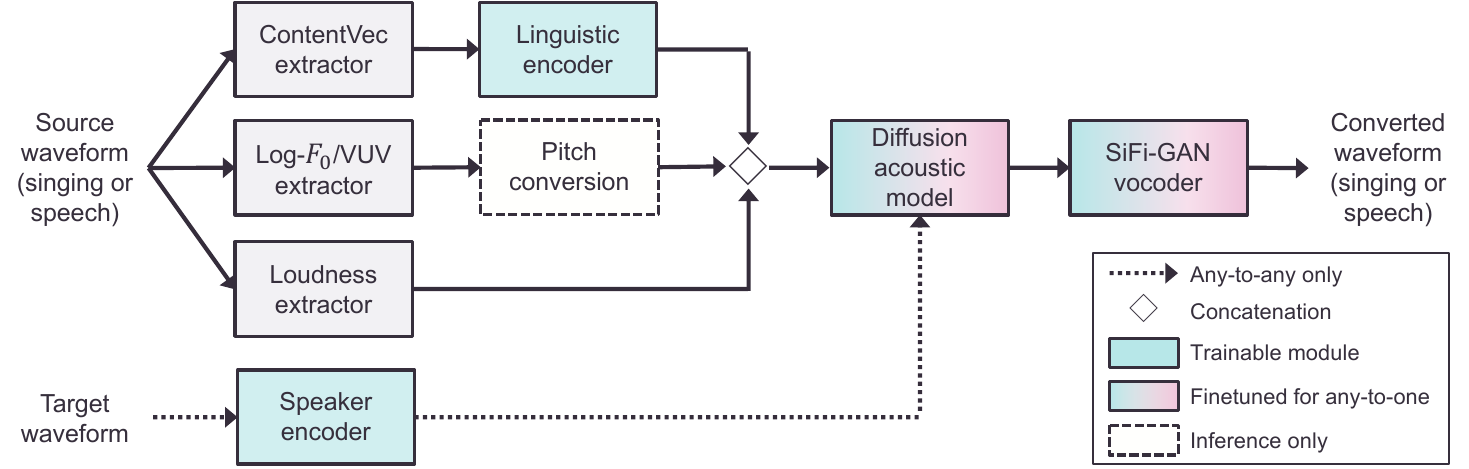,width=140mm}}
  \vspace{-2mm}
  \caption{\small An illustration of our any-to-any speech/singing voice conversion framework. We use a pre-trained ContentVec as a fixed feature extractor.
  The speaker encoder and the linguistic encoder are jointly trained with the diffusion-based acoustic model. The acoustic model and the vocoder are pre-trained on a large dataset and then fine-tuned for each target singer/speaker.}
  \label{fig:overview}
  \vspace{-4mm}
\end{figure*}

\redit{SVCC 2023 consists of two any-to-one SVC tasks: in-domain SVC (Task~1) and cross-domain SVC (Task~2)~\cite{huang2023singing}.
The organizers provide the target samples as the training data: target singer's \textit{singing data} for Task~1 and target speaker's \textit{speech data} for Task~2.
For both tasks, male and female target singers/speakers are provided: IDM1 and IDF1 for the in-domain task and CDM1 and CDF1 for the cross-domain task.
The goal for the participants is to develop better SVC systems that convert \textit{unknown} source singing to that of the target singers/speakers in terms of naturalness and speaker similarity.
The second task is considered harder since the singing data is not available for the target \redit{speakers}.
Note that the participants are allowed to use other publicly available datasets as additional training data.}

To address the problem of the limited amount of provided dataset, we utilize publicly available speech and singing datasets.
\redit{In total, we collect 750 hours of data that includes more than 2,700 speakers, comprising 630 hours of speech and 120 hours of singing data}. Using this large-scale dataset, we train a universal any-to-any VC model based on a recognition-synthesis framework~\cite{huang2022comparative}. \redit{Subsequently,} we finetune the universal VC model for the any-to-one \redit{SVC} cases of the two tasks in SVCC 2023.

As the VC model, we employ a strong diffusion probabilistic model for mel-spectrogram prediction to make the model learn the diverse characteristics of speech and singing~\cite{ho2020denoising}.
Instead of using phonetic posteriogram (PPG) or bottleneck features as the linguistic content features~\cite{liu2021diffsvc}, we adopt ContentVec-based features obtained by a self-supervised learning (SSL) with explicit speaker disentanglement~\cite{qian2022contentvec}.
This approach enables us to train the model on untranscribed datasets without relying on phonetic transcriptions.
To further disentangle speaker information from ContentVec features, we use a recently proposed information perturbation technique that makes our VC models generalize better~\cite{choi2021neural}.
We adopt the source-filter HiFi-GAN (SiFi-GAN) as a neural vocoder for high-fidelity SVC while achieving robustness for \redit{fundamental frequency ($F_0$) not in the training data}~\cite{yoneyama2023source}.
\redit{Although our framework is similar to the previous works using SSL representations for SVC~\cite{wang22u_interspeech, jayashankar2023self}, we aim to provide a comparative study of various training data configurations and model types for both in-domain and cross-domain SVC, which have not yet been well studied.}

\vspace{-2mm}
\section{Detail Description of our T13 System}
\vspace{-1mm}
\label{sec:method}


Figure \ref{fig:overview} shows the overview of our proposed VC/SVC framework.
Our method is a recognition-synthesis system with the following intermediate features:
\begin{inparaenum}[(1)]
\item linguistic content features based on ContentVec~\cite{qian2022contentvec},
\item logarithmic $F_0$ (log-$F_0$) and voiced/unvoiced flags (VUV),
\item loudness, and
\item speaker identity features (i.e., speaker embedding).
\end{inparaenum}
The VC/SVC process can be performed by first converting the speaker embedding to that of the target speaker and then using a synthesizer \redit{(i.e., acoustic model and vocoder)} to generate the target speech/singing.


\vspace{-2mm}
\subsection{Feature extraction}

To extract linguistic content features from the input waveform, we use ContentVec: an improved SSL representation with speaker disentanglement~\cite{qian2022contentvec}.
The model structure is the same as the HuBERT~\cite{hsu2021hubert}, but adopts a speaker disentanglement mechanism to learn a speaker-invariant representation without a significant loss of linguistic content.
We used a pre-trained model from the official GitHub repository~\footnote{\url{https://github.com/auspicious3000/contentvec}} as a fixed feature extractor, which was trained on 960 hours of LibriSpeech dataset~\cite{panayotov2015librispeech}.

For extracting log-$F_0$ and VUV, we use Harvest~\cite{harvest} and D4C (Definitive Decomposition Derived Dirt-Cheap)~\cite{morise2016d4c}, respectively.
A-weighting mechanism of a signal's power spectrum is used to compute the loudness features~\cite{liu2021diffsvc}.

Speaker embedding is extracted by a speaker encoder network from 80-dimensional log-scale mel-spectrogram.
The speaker encoder is based on a reference encoder with global style tokens (GSTs)~\cite{skerry2018towards,wang2018style}.

\vspace{-2mm}
\subsection{Acoustic model}
\vspace{-1mm}
\subsubsection{Mel-spectrogram prediction based on a denoising diffusion probabilistic model}

As an acoustic model, we employ a strong generative model based on a denoising diffusion probabilistic model~\cite{ho2020denoising} .
Similar to the SVCC 2023's baseline system (B01)~\cite{liu2021diffsvc}, we use a diffusion model to predict mel-spectrogram from the SSL features, log-$F_0$/VUV, and loudness features.
\redit{However, instead of PPG, we adopt ContentVec-based SSL features to enable us to utilize untranscribed datasets}.
Furthermore, we use speaker embeddings extracted from a speaker encoder network as the additional input.
To allow parameter-efficient fine-tuning for any-to-one SVC, we adopt conditional layer normalization~\cite{chen2021adaspeech} and make the diffusion model conditioned on the speaker embedding.
We use classifier-free guidance for better speaker adaptation
\footnote{
In our preliminary experiments, we confirmed it is possible to tradeoff speaker similarity and naturalness by controlling the guidance scale. However, increasing the guidance scale sometimes caused audible artifacts. We set the guidance scale to 1.0 for our submitted \nushort system.
}~\cite{ho2022classifier,kim2022guided}.

\vspace{-2mm}
\subsubsection{Linguistic encoder with information perturbation}
\label{sssec:ip}

Although SSL features can be used as linguistic features, previous studies suggest that SSL features contain speaker information that may degrade the VC performance~\cite{huang2022comparative,siuzdak2022wavthruvec}.
To address this issue, we apply an information perturbation technique to explicitly disentangle speaker information from the learned SSL features~\cite{choi2021neural}.

Specifically, we introduce a linguistic encoder network that processes the SSL features to the speaker-invariant linguistic features.
During training, two random perturbation functions \redit{that do not change the linguistic contents} are applied to the input waveform.
Then, \redit{a pair of SSL representations are extracted from the perturbed waveforms}.
Finally, the linguistic encoder network is trained to extract the same linguistic contents for those perturbed SSL representations with a contrastive loss.
We use formant shift, pitch randomization, and parametric equalizer as the perturbation methods~\cite{choi2022nansy++}.

\vspace{-2mm}
\subsection{Waveform generation by a source-filter neural vocoder}
\label{ssec:vocoder}

To synthesize the output waveform, we use SiFi-GAN~\cite{yoneyama2023source}, a source-filter neural vocoder based on HiFi-GAN~\cite{kong2020hifi}.
Thanks to the explicit $F_0$-driven architecture with a source-filter mechanism, SiFi-GAN can achieve high-fidelity waveform synthesis with \redit{robustness for $F_0$ values not present in the training data}.
This robustness is particularly beneficial for cross-domain SVC, where the target singer's pitch range is quite different from that of speech~\cite{duan2013nus}.

%

\begin{table*}[!t]
    \renewcommand{\arraystretch}{1.0}
        \caption{\small List of datasets used for training VC systems.
        Four different sets of training data are created to investigate the impact of large-scale datasets.
        The last column represents the databases used for our final SVCC submission.
        }
    \centering
    \small
    \scalebox{0.79}{
    \begin{tabular}{c|cccc|cccc}
    \toprule
    \multicolumn{5}{c}{{Dataset}} & \multicolumn{4}{c}{Training data}\\
    \midrule
    Name & Language & Type & \# Speakers & Hours & v1\_sing\_en & v2\_ssmix\_en & v3\_sing\_langmix & final \\
    \midrule
    VCTK~\cite{veaux2017cstr}          & en & speech & 109 & 41.03 & & \checkmark & & \checkmark \\
    LibriTTS~\cite{zen2019libritts}          & en & speech & 2456 & 585.83  & & \checkmark & & \checkmark \\
    NUS-48-E (speech)~\cite{duan2013nus}          & en & speech & 12 & 0.72  & & \checkmark & & \checkmark \\
    NUS-48-E (singing)~\cite{duan2013nus}         & en & singing & 12 & 1.55  & \checkmark & \checkmark & \checkmark & \checkmark \\
    Opencpop~\cite{wang2022opencpop}         & zh & singing & 1 & 5.23  & & & \checkmark & \checkmark \\
    OpenSinger~\cite{huang2021multi}         & zh & singing & 66 & 51.93  & & & \checkmark & \checkmark \\
    M4Singer~\cite{zhang2022m4singer}        & zh & singing & 20 & 29.7  & & & \checkmark & \checkmark \\
    PopCS~\cite{liu2021diffsinger}      & zh & singing & 1 & 5.89  & & & \checkmark & \checkmark \\
    CSD (en)~\cite{choi2020children}        & en & singing & 1 & 2.07  &\checkmark & \checkmark & \checkmark & \checkmark \\
    CSD (kr)~\cite{choi2020children}          & kr & singing & 1 & 2.23  & & & \checkmark & \checkmark \\
    KSing~\cite{ksing}         & zh & singing & 1 & 0.89  & & & \checkmark & \checkmark \\
    JSUT song~\cite{sonobe2017jsut}      & ja & singing & 1 & 0.37  & & & \checkmark & \checkmark \\
    Tohoku Kiritan~\cite{ogawa2021tohoku}      & ja & singing & 1 & 1.07  & & & \checkmark & \checkmark \\
    JVS-MuSIC~\cite{tamaru2020jvs}      & ja & singing & 100 & 3.59  & & & \checkmark & \checkmark \\
    Misc. Japanese singing DBs    & ja & singing & 8 & 17.45  & & & \checkmark & \checkmark \\
    \midrule
    SVCC 2023 (subset of NHSS~\cite{sharma2021nhss})      & en & \redit{speech/singing} & 4 & 0.59  & \checkmark & \checkmark & \checkmark & \checkmark \\
     \bottomrule
    \end{tabular}
    }
    \label{tab:dataset}
    \vspace{-4mm}
    \renewcommand{\arraystretch}{1.0}
\end{table*}

\vspace{-2mm}
\subsection{Training}
\label{ssec:training}

The training process is divided into two stages: pre-training on large-scale data and fine-tuning for a specific singer/speaker of in-domain and cross-domain SVC tasks.

During pre-training, the speaker encoder, the linguistic encoder, and the acoustic models are jointly trained on a large-scale speech/singing dataset.
The loss function is a combination of an L2 loss for the diffusion-based acoustic model and a contrastive loss for information perturbation.
\redit{We linearly increase the weight for the contrastive loss by $1e^{-5} \times n$, where $n$ represents the training step~\cite{qian2022contentvec}.}
For the vocoder, we simply train a universal SiFi-GAN on the same dataset with reconstruction and adversarial objectives~\cite{yoneyama2023source}.

After pre-training, we finetune the pre-trained acoustic model for each singer/speaker of the given SVCC dataset.
Due to the any-to-one settings of the SVCC tasks, the speaker encoder is not used for the fine-tuning process.
Instead, we \redit{use} a fixed pseudo speaker embedding that is normalized to have a unit norm~\cite{kim2022guided}.
Note that information perturbation and contrastive loss are not used for fine-tuning.
The pre-trained universal SiFi-GAN vocoder is also fine-tuned \rredit{using the ground-truth mel-spectrogram extracted from} the given dataset to further improve the performance.


\vspace{-2mm}
\subsection{Pitch conversion}

To convert source speech or singing to target one,
the source waveform is first decomposed into linguistic content, log-$F_0$/VUV, and loudness features.
To make the converted pitch sounds like the target, we adopt simple mean-variance normalization of the log-$F_0$ as follows:
\begin{align}
\hat{f}_{t} = \frac{\sigma^{(y)}}{\sigma^{(x)}} (f_{t} - \mu^{(x)}) + \mu^{(y)}
\end{align}
where $f_{t}$ and $\hat{f_{t}}$ denote the log-$F_0$ of the source speaker and converted one at frame $t$, $\mu^{(x)}$ and $\mu^{(y)}$ denote the mean of the log-$F_0$ for the source and target speakers, and $\sigma^{(x)}$ and $\sigma^{(y)}$ denote the standard deviation of the log-$F_0$ for the source and target speakers, respectively.
The mean and standard deviations of log-$F_0$ are computed from the training data~\footnote{
For the SVCC's source singers, we computed the statistics using the evaluation dataset since it was impossible to estimate the statistics of singers not in the training data.
}.

To further improve the naturalness of the converted pitch, we use the following heuristics for SVCC: \begin{inparaenum}[(1)]
\item $\sigma^{(x)}$ and $\sigma^{(y)}$ are set to one; performing pitch-shift only for singing to avoid out-of-tune pitch.
\item The amount of pitch shift ($\mu^{(y)} - \mu^{(x)}$) is quantized in 100~cents. 
\item We increase the log-$F_0$ with \rredit{six semitones} for cross-domain SVC only.
Note that the value \rredit{six} was chosen based on the statistics of the NUS-48-E: a parallel speech/singing dataset~\cite{duan2013nus}.
\end{inparaenum}



\section{Experimental evaluations}
\label{sec:exp}

\subsection{Datasets}


Table \ref{tab:dataset} summarizes the datasets used for training our models.
In addition to the provided SVCC 2023 dataset, we collected publicly available singing and speech datasets with high-quality audio of sampling rates higher than 24~kHz.
All the audio files were resampled to 24~kHz.
Because some of the singing datasets contain unsegmented long audio files, we performed automatic segmentation based on the rest note information if the musical score is available, otherwise we used voice activity detection-based segmentation~\footnote{\url{https://github.com/wiseman/py-webrtcvad}}.
In total, we used 750 hours of segmented data containing approximately 500~K audio clips.
Note that although the most datasets contain lyrics or text transcriptions, we did not use them to allow our model scale for untranscribed datasets.

To investigate the impact of mixing a large number of datasets, we perform experiments with the following four sets of training data.
\noindent\textbf{v1\_sing\_en}: English only singing datasets (4 hours) \\
\noindent\textbf{v2\_ssmix\_en}: English only speech and singing datasets (630 hours)\\
\noindent\textbf{v3\_sing\_langmix}: Mixed language singing datasets (120 hours)\\
\noindent\textbf{final}: All the datasets (750 hours)\\
\redit{We included the target singers/speakers (i.e., IDF1, IDM1, CDF1, and CDM1) in all the training sets.}



\begin{table}[tb]
\vspace{-2.5mm}
\caption{\small \redit{Number of parameters of our \nushort system}}
\label{tab:model_params}
\centering
\scalebox{0.85}{
\begin{tabular}{l|c}
\toprule
Module & \# Parameters (million) \\
\midrule
ContentVec & 94.6 \\
Linguistic encoder & 1.3 \\
Speaker encoder & 5.8 \\
Diffusion-based acoustic model & 133 \\
SiFi-GAN vocoder &  102 \\
\bottomrule
\end{tabular}
}
\vspace{-6mm}
\end{table}

\subsection{Model details}

\redit{Table~\ref{tab:model_params} shows the number of parameters of our \redit{submitted} \nushort system}.
The \rredit{diffusion acoustic model uses a denoiser based on} a simplified non-causal WaveNet~\cite{liu2021diffsinger}.
The model contains 20-layers of non-causal residual one-dimensional convolution layers with skip connections.
To investigate the impact of the model size, we used two different models with the channel sizes of 256 and 768 for the base and large models, respectively.
We used the large model for our final submission, but used the base model to compare different training data and model configurations.
The number of diffusion steps was set to 100.
We trained the acoustic models for 100 epochs (\redit{670~K steps for the final dataset}) for pre-training.
For fine-tuning, we updated the parameters of the conditional layer normalization modules for 500 iterations~\cite{chen2021adaspeech}.
We used AdamW optimizer~\cite{loshchilov2017decoupled} with a batch size of 4~K frames.
Pre-training took approximiately 8 days using a single Tesla A100 GPU.

The 768-dimensional ContentVec features were converted to 128-dimensional linguistic features by the linguistic encoder.
We used the hidden features of ContentVec before the final projection layer.
The linguistic encoder consists of six-layers one-dimensional convolution layers with residual connections.
The channel sizes of the convolution layers were set to 128.

\redit{For our GST-based speaker encoder}~\cite{wang2018style}, we used 128, 128, 256, 256, 512 and 512 output channels for \redit{six} convolutional layers, respectively.
The number of hidden units in a gated recurrent unit was set to 256.
We set the number of style tokens, their dimensionality, and the number of attention heads to 50, 256, and 4, respectively.

As the vocoder, a universal SiFi-GAN model was trained on the final dataset \redit{(i.e. 750 hours of speech and singing)} for 2,000~K steps.
To enhance the generalization ability of the SiFi-GAN vocoder, we set the channel size of convolution layers to 1536, which is \redit{3 times larger} than the settings in the original SiFi-GAN\footnote{
As suggested in prior work on universal vocoders~\cite{lee2022bigvgan}, we confirmed that larger vocoders worked better when trained on a large dataset.
}~\cite{yoneyama2023source}.
We used Adam optimizer~\cite{kingma2014adam} for training the vocoder.
Pre-training took about two weeks using a single Tesla A100 GPU.
We performed fine-tuning for 20~K iterations.

To investigate the effectiveness of ContentVec, we compared SVC models with HuBERT-soft as the content features~\cite{van2022comparison}.
The number of hidden features of HuBERT-soft was 256.
We also compare SVC models without information perturbation to confirm the effectiveness of the speaker disentanglement.
The linguistic encoder \redit{was} omitted for the VC models without information perturbation and SSL features were directly fed to the acoustic model.

\subsection{Objective evaluation}

\begin{table*}[!t]
    \vspace{-2.5mm}
    \renewcommand{\arraystretch}{1.0}
        \caption{\small Task~1: In-domain SVC results for the SVCC evaluation dataset.
        Our \nushort system is denoted as S8.
        AM and IP represent the acoustic model and information perturbation, respectively.
        SOU represents the source recorded samples.
        Note that target samples are omitted since they are not provided by the challenge.
        }
    \centering
    \small
    \scalebox{0.83}{
    \begin{tabular}{l| cccc | ccc | ccc }
    \toprule
    \multicolumn{5}{c}{{Model}} & \multicolumn{3}{c}{Pre-training} & \multicolumn{3}{c}{Fine-tuning} \\
    \midrule
    System & Training data & SSL & AM & IP & UTMOS~($\uparrow$) & COSSIM~($\uparrow$)  & WER~($\downarrow$)  & UTMOS~($\uparrow$)  & COSSIM~($\uparrow$)  & WER~($\downarrow$)  \\
     \midrule
    S1 & v1\_sing\_en & ContentVec & Base &  & 1.969 & 0.797 & 24.0 & 1.947 & 0.801 & 24.3 \\
    S2 & v2\_ssmix\_en & ContentVec & Base &  & 2.038 & 0.825 & 17.1 & 2.090 & 0.826 & 18.7 \\
    S3 & v3\_sing\_langmix & ContentVec & Base &  & \textbf{2.169} & 0.826 & \textbf{15.7} & 2.127 & 0.831 & 16.5 \\
    \midrule
    S4 & final & HuBERT-soft & Base &            & 2.128 & 0.829 & 21.8 & 2.137 & 0.810 & 19.1 \\
    S5 & final & HuBERT-soft & Base & \checkmark & 2.151 & \textbf{0.839} & 34.1 & 2.179 & 0.821 & 26.7 \\
    S6 & final & ContentVec & Base &            & 2.154 & 0.829 & 16.4 & 2.189 & 0.822 & \textbf{16.2} \\
    S7 & final & ContentVec & Base & \checkmark & 2.113 & 0.833 & 23.3 & 2.183 & 0.829 & 19.1\\
    S8 & final & ContentVec & Large & \checkmark & 2.162 & 0.835 & 26.9 & \textbf{2.225} & \textbf{0.834} & 23.2 \\
    \midrule
    SOU & - & - & - & - & 2.167 & - & 7.3 & 2.167 & - & 7.3 \\
    \bottomrule
    \end{tabular}
    }
    \label{tab:eval_svcc1}
    \vspace{-3mm}
    \renewcommand{\arraystretch}{1.0}
\end{table*}

\begin{table*}[!t]
    \vspace{-2mm}
    \renewcommand{\arraystretch}{1.0}
        \caption{\small Task~2: Cross-domain SVC results for the SVCC evaluation dataset.
        }
    \vspace{1mm}
    \centering
    \small
    \scalebox{0.83}{
    \begin{tabular}{l| cccc | ccc | ccc }
    \toprule
    \multicolumn{5}{c}{{Model}} & \multicolumn{3}{c}{Pre-training} & \multicolumn{3}{c}{Fine-tuning} \\
    \midrule
    System & Training data & SSL & AM & IP & UTMOS~($\uparrow$) & COSSIM~($\uparrow$)  & WER~($\downarrow$)  & UTMOS~($\uparrow$)  & COSSIM~($\uparrow$)  & WER~($\downarrow$)  \\
     \midrule
    S1 & v1\_sing\_en & ContentVec & Base &  & 2.010 & 0.758 & 26.2 & 2.002 & 0.774 & 24.2 \\
    S2 & v2\_ssmix\_en & ContentVec & Base &  & 2.300 & 0.804 & \textbf{16.0} & 2.308 & 0.828 & \textbf{16.4} \\
    S3 & v3\_sing\_langmix & ContentVec & Base &  & 2.383 & 0.818 & 20.0 & 2.314 & 0.828 & 17.1 \\
    \midrule
    S4 & final & HuBERT-soft & Base &            & 2.342 & 0.813 & 21.9 & 2.333 & 0.810 & 21.8 \\
    S5 & final & HuBERT-soft & Base & \checkmark & 2.393 & 0.817 & 30.2 & 2.397 & 0.828 & 29.4 \\
    S6 & final & ContentVec & Base &            & 2.357 & 0.814 & 17.5 & 2.387 & 0.826 & 17.1 \\
    S7 & final & ContentVec & Base & \checkmark & 2.339 & 0.817 & 25.4 & 2.393 & 0.838 & 20.0 \\
    S8 & final & ContentVec & Large & \checkmark & \textbf{2.398} & \textbf{0.824} & 23.6 & \textbf{2.456} & \textbf{0.842} & 20.4 \\
    \midrule
    SOU & - & - & - & - & 2.167 & - & 7.3 & 2.167 & - & 7.3 \\
    \bottomrule
    \end{tabular}
    }
    \label{tab:eval_svcc2}
    \vspace{-3mm}
    \renewcommand{\arraystretch}{1.0}
\end{table*}

\begin{figure}[!tb]
\begin{center}
    \vspace{-2.0mm}
    \includegraphics[width=0.78\columnwidth]{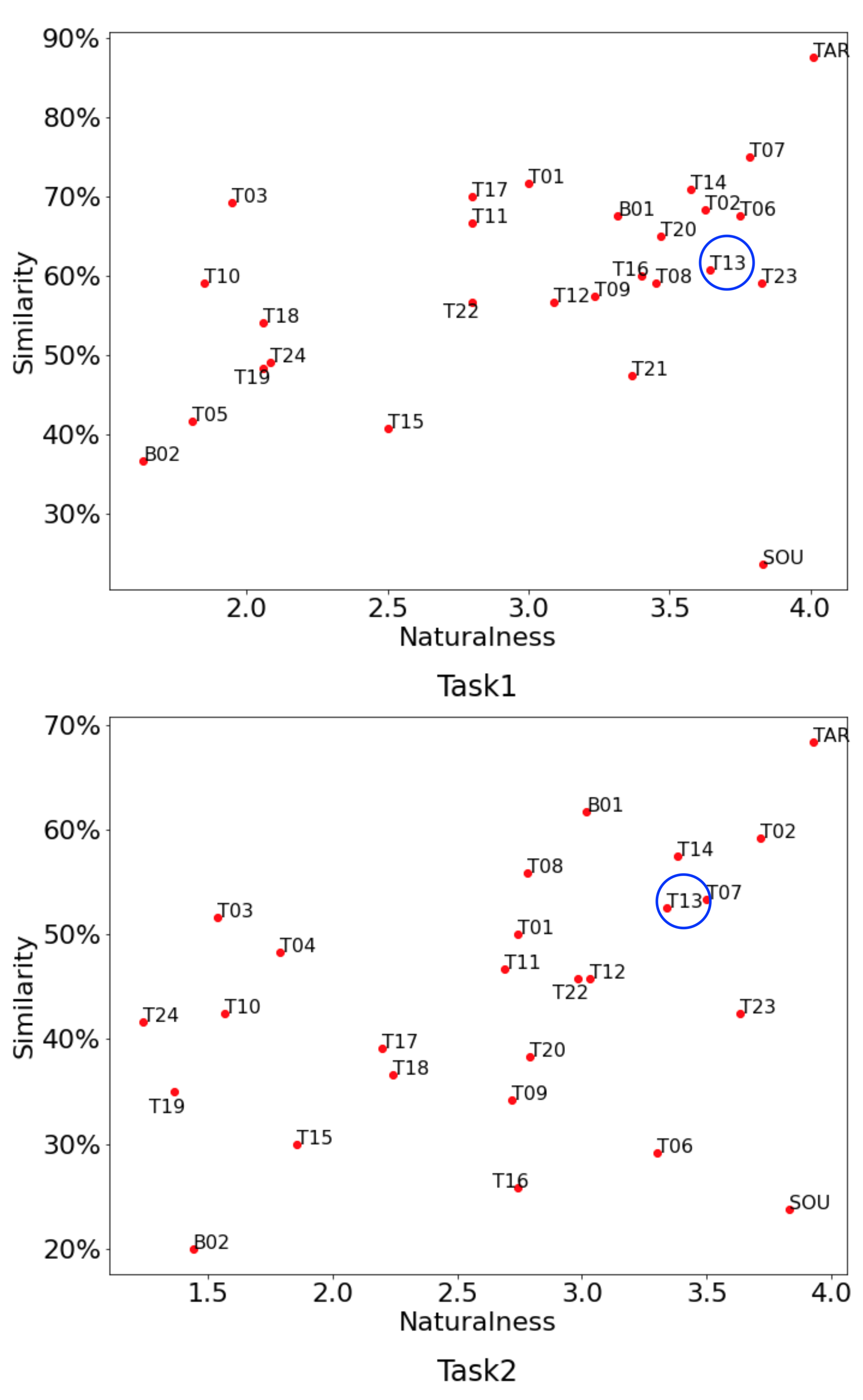}
    \vspace{-4mm}
    \caption{
    \small Scatter plots of naturalness and similarity percentage for Task 1 (in-domain) and Task 2 (cross-domain) from English listeners~\cite{huang2023singing}. \redit{Our system is denoted as T13 with blue circles.}
    }
    \label{fig:svcc2023sub}
    \end{center}
\vspace{-10mm}
\end{figure}

We conducted experiments with two task configurations:
\begin{inparaenum}[(1)]
\item Task~1: in-domain SVC
\item Task~2: cross-domain SVC.
\end{inparaenum}
We used the two target singers/speakers for each task: IDM1/IDF1 for Task~1, and CDM1/CDF1 for Task~2, respectively.
\redit{As the source singers, one male and one female singers were used from the SVCC evaluation dataset.
We tested four pairs of SVC: male-to-male, male-to-female, female-to-male, and female-to-female.}
\redit{We used 24 source samples for each source singer.}
\redit{In total, 48 samples were generated for each target singer/speaker}.


\noindent\textbf{Objective metrics}:
We used UTMOS as a naturalness mean opinion score (MOS) predictor~\cite{saeki2022utmos}.
MOS was estimated for each utterance, and then we took the average MOS values for each model type.
To measure speaker similarity, we computed the cosine similarity between speaker embeddings of source and target samples.
We used a pre-trained WavLM-based speaker verification model~\footnote{
\url{https://huggingface.co/microsoft/wavlm-base-plus-sv}
} for extracting speaker embeddings~\cite{chen2022wavlm}. 
We computed the average cosine similarity (COSSIM) between the source and randomly selected target samples from the SVCC dataset.
For Task~2, we used the target speech as the reference for computing the similarity since the target singing is not available.
To evaluate intelligibility, we measured word error rate (WER) using a robust speech recognition system based on Whisper (large-V2)~\cite{radford2022robust}.
The beam size for decoding was set to 15. %


\vspace{-1mm}
\subsubsection{Task~1: in-domain SVC}

Table~\ref{tab:eval_svcc1} shows the objective evaluation results for in-domain SVC.
The findings are summarized as follows.
\begin{inparaenum}[(1)]
\item The models trained on large speech or singing datasets outperformed the model trained on the small dataset in all the metrics (S1~vs.~S2; S1~vs.~S3).
\item The method using large singing datasets outperformed the method using large speech datasets (S2~vs.~S3), implying that using large singing datasets is more beneficial than using large speech datasets. \rredit{Additionally, no significant negative effects were observed from mixing multiple languages.}
\item Information perturbation improved the speaker similarity for both ContentVec and HuBERT-soft-based methods, whereas certain degradation in intelligibility was observed.
\item \redit{The models using ContentVec outperformed the models with HuBERT-soft features in intelligibility. Furthermore, HuBERT-soft-based methods suffered more from speaker similarity degradation with fine-tuning, possibly due to the insufficient speaker disentanglement~\cite{huang2022comparative}.}
\item \redit{Fine-tuning improved the intelligibility of the models trained on the final dataset. This result implies that learning the diversity of pronunciations in speech and singing was challenging. Thus, fine-tuning was necessary to maximize the performance of the pre-trained models.}
\item \redit{All the systems trained on the final dataset achieved comparable performance, suggesting that data size matters more than the model configurations (e.g., model size).}
\end{inparaenum}



\vspace{-1mm}
\subsubsection{Task~2: cross-domain SVC}

Table~\ref{tab:eval_svcc2} shows the objective evaluation results for cross-domain SVC.
Similar trends can be observed compared to the results of Task~1.
\redit{However, compared to the results of Task~1, we found that using the large speech and singing datasets contributed more to improving the SVC performance (S1~vs.~S2; S1~vs.~S3).
For example, when comparing \redit{fine-tuned} S1 and S2, a speaker similarity improvement of \rredit{0.054} was obtained for Task 2, while that of Task 1 was only \rredit{0.025}.
The same tendency was observed for the naturalness scores.}
\redit{These results imply that acquiring a general representation from the large speech and singing datasets effectively enabled the model to generalize well in the more challenging cross-domain SVC scenarios.
Our submitted system (S8) performed the best regarding naturalness and speaker similarity, but the intelligibility was worse than the method trained without information perturbation (S6).}

Note that we observed that the naturalness of the converted singing voice was often higher than that of the recorded source singing.
This can be attributed to the fact that UTMOS tends to assign higher scores to samples that resemble speech, as the model was trained on speech datasets.
Although a moderate correlation exists between UTMOS \redit{scores and perceived naturalness}~\cite{huang2023singing}, predicting naturalness specifically for singing remains an important direction for future research.

\subsection{Subjective evaluations}

To evaluate the perceptual naturalness and speaker similarity, large-scale listening tests were performed by SVCC 2023.
Details can be found in the SVCC 2023 paper~\cite{huang2023singing}.

Figure~\ref{fig:svcc2023sub} shows the listening test results from English raters.
Our system is denoted as T13, which corresponds to the fine-tuned S8 in Table~\ref{tab:eval_svcc1} and Table~\ref{tab:eval_svcc2}.
For Task~1, our system \rredit{achieved} relatively high naturalness, but the speaker similarity was average among all the submitted systems.
\redit{We hypothesize that the average speaker similarity is because most of our training data consists of speech data. As a result, the trained model was biased towards generating samples closer to speech.}
On the other hand, our system achieved competitive scores for both the naturalness and speaker similarity metrics in the more challenging Task~2: T13 is located in the top right area in the scatter plot. These results demonstrate the generalization capability of the proposed method.

\redit{We found that the models trained on large datasets can generalize well for any-to-any scenarios.
We encourage readers to listen to the audio samples at our demo page~\footnotemark[1].}

\section{Conclusion}
\label{sec:conclusion}

This paper presented our systems (T13) for the singing voice conversion challenge 2023.
We adopt a recognition-synthesis approach with ContentVec features and an additional linguistic encoder.
To address low-resource issues of SVC, we first train a diffusion-based any-to-any VC model using publicly available large-scale 750 hours of speech and singing data.
Then, we finetune the model for each target singer/speaker of Task~1 and Task~2.
Experimental results showed that our \nushort system achieved competitive naturalness and speaker similarity for the harder cross-domain SVC (Task~2).
The objective evaluation results showed that using the large-scale dataset was particularly helpful for cross-domain SVC.
\redit{Future work includes investigating SVC models for more challenging any-to-any in-domain/cross-domain SVC. Additionally, exploring the relationship between data size and generalization ability is worthwhile.}

\vspace{-1mm}
\hfill \break
\noindent\textbf{Acknowledgments}:
\rredit{This work was supported in part by JST CREST Grant Number JPMJCR19A3, Japan, and JSPS KAKENHI Grant Number 21H05054.}



\section{References}
{
\setstretch{0.89}
\printbibliography

@article{liu2021diffsinger,
  title={{DiffSinger: Singing voice synthesis via shallow diffusion mechanism}},
  author={Liu, Jinglin and Li, Chengxi and Ren, Yi and Chen, Feiyang and Liu, Peng and Zhao, Zhou},
  journal={AAAI},
  volume={36},
  number={10},
  pages={11020--11028},
  year={2022}
}

@article{sharma2021nhss,
  title={{NHSS}: A speech and singing parallel database},
  author={Sharma, Bidisha and Gao, Xiaoxue and Vijayan, Karthika and Tian, Xiaohai and Li, Haizhou},
  journal={Speech Communication},
  volume={133},
  pages={9--22},
  year={2021},
  publisher={Elsevier}
}

@inproceedings{liu2021diffsvc,
  title={{DiffSVC}: A diffusion probabilistic model for singing voice conversion},
  author={Liu, Songxiang and Cao, Yuewen and Su, Dan and Meng, Helen},
  booktitle={Proc. ASRU},
  pages={741--748},
  year={2021},
  organization={IEEE}
}

@inproceedings{wang2022opencpop,
  author={Yu Wang and Xinsheng Wang and Pengcheng Zhu and Jie Wu and Hanzhao Li and Heyang Xue and Yongmao Zhang and Lei Xie and Mengxiao Bi},
  title={{Opencpop: A High-Quality Open Source Chinese Popular Song Corpus for Singing Voice Synthesis}},
  year=2022,
  booktitle={Proc. Interspeech 2022},
  pages={4242--4246},
}

@inproceedings{huang2021multi,
  title={Multi-singer: Fast multi-singer singing voice vocoder with a large-scale corpus},
  author={Huang, Rongjie and Chen, Feiyang and Ren, Yi and Liu, Jinglin and Cui, Chenye and Zhao, Zhou},
  booktitle={Proc. ACM ICM},
  pages={3945--3954},
  year={2021}
}

@article{zhang2022m4singer,
  title={M4Singer: A Multi-Style, Multi-Singer and Musical Score Provided Mandarin Singing Corpus},
  author={Zhang, Lichao and Li, Ruiqi and Wang, Shoutong and Deng, Liqun and Liu, Jinglin and Ren, Yi and He, Jinzheng and Huang, Rongjie and Zhu, Jieming and Chen, Xiao and others},
  journal={Proc. NeurIPS},
  volume={35},
  pages={6914--6926},
  year={2022}
}

@inproceedings{choi2020children,
  title={Children’s song dataset for singing voice research},
  author={Choi, Soonbeom and Kim, Wonil and Park, Saebyul and Yong, Sangeon and Nam, Juhan},
  booktitle={International Society for Music Information Retrieval Conference (ISMIR)},
  year={2020}
}

@article{tamaru2020jvs,
  title={{JVS-MuSiC}: Japanese multispeaker singing-voice corpus},
  author={Tamaru, Hiroki and Takamichi, Shinnosuke and Tanji, Naoko and Saruwatari, Hiroshi},
  journal={arXiv preprint arXiv:2001.07044},
  year={2020}
}

@inproceedings{koguchi2020pjs,
  title={{PJS}: Phoneme-balanced Japanese singing-voice corpus},
  author={Koguchi, Junya and Takamichi, Shinnosuke and Morise, Masanori},
  booktitle={Proc. APSIPA ASC},
  pages={487--491},
  year={2020},
  organization={IEEE}
}

@article{ogawa2021tohoku,
  title={Tohoku Kiritan singing database: A singing database for statistical parametric singing synthesis using Japanese pop songs},
  author={Ogawa, Itsuki and Morise, Masanori},
  journal={Acoustical Science and Technology},
  volume={42},
  number={3},
  pages={140--145},
  year={2021},
  publisher={Acoustical Society of Japan}
}

@inproceedings{yoneyama2023source,
  title={{Source-Filter HiFi-GAN: Fast and Pitch Controllable High-Fidelity Neural Vocoder}},
  author={Yoneyama, Reo and Wu, Yi-Chiao and Toda, Tomoki},
  booktitle={ICASSP 2023-2023 IEEE International Conference on Acoustics, Speech and Signal Processing (ICASSP)},
  pages={1--5},
  year={2023},
  organization={IEEE}
}

@inproceedings{van2022comparison,
  title={A comparison of discrete and soft speech units for improved voice conversion},
  % author={van Niekerk, Benjamin and Carbonneau, Marc-Andr{\'e} and Za{\"\i}di, Julian and Baas, Matthew and Seut{\'e}, Hugo and Kamper, Herman},
  % TODO
  author={van Niekerk, Benjamin and Carbonneau, et al.},
  booktitle={ICASSP 2022-2022 IEEE International Conference on Acoustics, Speech and Signal Processing (ICASSP)},
  pages={6562--6566},
  year={2022},
  organization={IEEE}
}

@article{huang2022comparative,
  title={A comparative study of self-supervised speech representation based voice conversion},
  author={Huang, Wen-Chin and Yang, Shu-Wen and Hayashi, Tomoki and Toda, Tomoki},
  journal={IEEE Journal of Selected Topics in Signal Processing},
  volume={16},
  number={6},
  pages={1308--1318},
  year={2022},
  publisher={IEEE}
}

@article{huang2023singing,
  title={The Singing Voice Conversion Challenge 2023},
  author={Huang, Wen-Chin and Violeta, Lester Phillip and Liu, Songxiang and Shi, Jiatong and Yasuda, Yusuke and Toda, Tomoki},
  journal={arXiv preprint arXiv:2306.14422},
  year={2023}
}

@inproceedings{liu2021fastsvc,
  title={{FastSVC}: Fast cross-domain singing voice conversion with feature-wise linear modulation},
  author={Liu, Songxiang and Cao, Yuewen and Hu, Na and Su, Dan and Meng, Helen},
  booktitle={Proc. ICME},
  pages={1--6},
  year={2021},
  organization={IEEE}
}

@inproceedings{duan2013nus,
  title={The {NUS} sung and spoken lyrics corpus: A quantitative comparison of singing and speech},
  author={Duan, Zhiyan and Fang, Haotian and Li, Bo and Sim, Khe Chai and Wang, Ye},
  booktitle={Proc. APSIPA ASC},
  pages={1--9},
  year={2013},
  organization={IEEE}
}

@inproceedings{jayashankar2023self,
  title={Self-Supervised Representations for Singing Voice Conversion},
  author={Jayashankar, Tejas and Wu, Jilong and Sari, Leda and Kant, David and Manohar, Vimal and He, Qing},
  booktitle={ICASSP 2023-2023 IEEE International Conference on Acoustics, Speech and Signal Processing (ICASSP)},
  pages={1--5},
  year={2023},
  organization={IEEE}
}

@misc{ksing,
  title = {{KiSing}: the First Open-source {Mandarin} Singing Voice Synthesis Corpus},
  author = {Jiatong Shi},
  howpublished = {\url{http://shijt.site/index.php/2021/05/16/}},
  note = {Accessed: 2023.07.16}
}

@inproceedings{wang22u_interspeech,
  author={Chao Wang and Zhonghao Li and Benlai Tang and Xiang Yin and Yuan Wan and Yibiao Yu and Zejun Ma},
  title={{Towards high-fidelity singing voice conversion with acoustic reference and contrastive predictive coding}},
  year=2022,
  booktitle={Proc. Interspeech 2022},
  pages={4287--4291},
}

@inproceedings{chen2019singing,
  title={Singing voice conversion with non-parallel data},
  author={Chen, Xin and Chu, Wei and Guo, Jinxi and Xu, Ning},
  booktitle={Proc. MIPR},
  pages={292--296},
  year={2019},
  organization={IEEE}
}

@inproceedings{polyak20b_interspeech,
  author={Adam Polyak and Lior Wolf and Yossi Adi and Yaniv Taigman},
  title={{Unsupervised Cross-Domain Singing Voice Conversion}},
  year=2020,
  booktitle={Proc. Interspeech 2020},
  pages={801--805},
}

@inproceedings{kingma2014adam,
  title={Adam: A method for stochastic optimization},
  author={Kingma, Diederik P and Ba, Jimmy},
  booktitle={Proc. ICLR},
  year={2015},
}

@inproceedings{wang2018style,
  title={Style tokens: Unsupervised style modeling, control and transfer in end-to-end speech synthesis},
  author={Wang, Yuxuan and Stanton, Daisy and Zhang, Yu and Ryan, RJ-Skerry and Battenberg, Eric and Shor, Joel and Xiao, Ying and Jia, Ye and Ren, Fei and Saurous, Rif A},
  booktitle={Proc. ICML},
  pages={5180--5189},
  year={2018},
  organization={PMLR}
}

@inproceedings{skerry2018towards,
  title={Towards end-to-end prosody transfer for expressive speech synthesis with {Tacotron}},
  author={Skerry-Ryan, RJ and Battenberg, Eric and Xiao, Ying and Wang, Yuxuan and Stanton, Daisy and Shor, Joel and Weiss, Ron and Clark, Rob and Saurous, Rif A},
  booktitle={Proc. ICML},
  pages={4693--4702},
  year={2018},
  organization={PMLR}
}

@inproceedings{kong2020hifi,
  title={{HiFi-GAN}: Generative adversarial networks for efficient and high fidelity speech synthesis},
  author={Kong, Jungil and Kim, Jaehyeon and Bae, Jaekyoung},
  booktitle={Proc. NeurIPS},
  volume={33},
  pages={17022--17033},
  year={2020}
}

@inproceedings{loshchilov2017decoupled,
  title={Decoupled weight decay regularization},
  author={Loshchilov, Ilya and Hutter, Frank},
  booktitle={Proc. ICLR},
  year={2019}
}

@article{ho2020denoising,
  title={Denoising diffusion probabilistic models},
  author={Ho, Jonathan and Jain, Ajay and Abbeel, Pieter},
  journal={Proc. NeurIPS},
  volume={33},
  pages={6840--6851},
  year={2020}
}

@article{veaux2017cstr,
  title={{CSTR VCTK corpus: English multi-speaker corpus for CSTR voice cloning toolkit}},
  author={Veaux, Christophe and Yamagishi, Junichi and MacDonald, Kirsten and others},
  journal={University of Edinburgh. The Centre for Speech Technology Research (CSTR)},
  year={2017},
  doi={10.7488/ds/2645}
}

@inproceedings{zen2019libritts,
  author={Heiga Zen and Viet Dang and Rob Clark and Yu Zhang and Ron J. Weiss and Ye Jia and Zhifeng Chen and Yonghui Wu},
  title={{LibriTTS: A Corpus Derived from {LibriSpeech} for Text-to-Speech}},
  year=2019,
  booktitle={Proc. Interspeech 2019},
  pages={1526--1530},
}

@inproceedings{zhao2020voice,
  title={Voice conversion challenge 2020: Intra-lingual semi-parallel and cross-lingual voice conversion},
  author={Zhao, Yi and Huang, Wen-Chin and Tian, Xiaohai and Yamagishi, Junichi and Das, Rohan Kumar and Kinnunen, Tomi and Ling, Zhenhua and Toda, Tomoki},
  booktitle={Proc. Joint Workshop for the BC and VCC},
  pages={80--98},
  year={2020}
}

@inproceedings{lorenzo2018voice,
  title={The voice conversion challenge 2018: Promoting development of parallel and nonparallel methods},
  author={Lorenzo-Trueba, Jaime and Yamagishi, Junichi and Toda, Tomoki and Saito, Daisuke and Villavicencio, Fernando and Kinnunen, Tomi and Ling, Zhenhua},
  booktitle={Proc. Odyssey},
  pages={195--202},
  year={2018}
}

@inproceedings{panayotov2015librispeech,
  title={{LibriSpeech}: an {ASR} corpus based on public domain audio books},
  author={Panayotov, Vassil and Chen, Guoguo and Povey, Daniel and Khudanpur, Sanjeev},
  booktitle={Proc. ICASSP},
  pages={5206--5210},
  year={2015},
  organization={IEEE}
}

@inproceedings{qian2022contentvec,
  title={{CONTENTVEC}: An improved self-supervised speech representation by disentangling speakers},
  author={Qian, Kaizhi and Zhang, Yang and Gao, Heting and Ni, Junrui and Lai, Cheng-I and Cox, David and Hasegawa-Johnson, Mark and Chang, Shiyu},
  booktitle={International Conference on Machine Learning},
  pages={18003--18017},
  year={2022},
  organization={PMLR}
}

@inproceedings{choi2022nansy++,
  title={{NANSY++}: Unified Voice Synthesis with Neural Analysis and Synthesis},
  author={Choi, Hyeong-Seok and Yang, Jinhyeok and Lee, Juheon and Kim, Hyeongju},
  booktitle={Proc. ICLR},
  year={2022},
}

@article{hsu2021hubert,
  title={{HuBERT}: Self-supervised speech representation learning by masked prediction of hidden units},
  author={Hsu, Wei-Ning and Bolte, Benjamin and Tsai, Yao-Hung Hubert and Lakhotia, Kushal and Salakhutdinov, Ruslan and Mohamed, Abdelrahman},
  journal={IEEE/ACM Trans. on Audio, Speech, and Lang. Process.},
  volume={29},
  pages={3451--3460},
  year={2021},
  publisher={IEEE}
}

@inproceedings{ho2022classifier,
  title={Classifier-free diffusion guidance},
  author={Ho, Jonathan and Salimans, Tim},
  booktitle={Proc. NeurIPS},
  year={2021},
}

@article{kim2022guided,
  title={{Guided-TTS 2}: A diffusion model for high-quality adaptive text-to-speech with untranscribed data},
  author={Kim, Sungwon and Kim, Heeseung and Yoon, Sungroh},
  journal={arXiv preprint arXiv:2205.15370},
  year={2022}
}

@inproceedings{chen2021adaspeech,
  title={Adaspeech: Adaptive text to speech for custom voice},
  author={Chen, Mingjian and Tan, Xu and Li, Bohan and Liu, Yanqing and Qin, Tao and Zhao, Sheng and Liu, Tie-Yan},
  booktitle={Proc. ICLR},
  year={2021},
}

@article{choi2021neural,
  title={Neural analysis and synthesis: Reconstructing speech from self-supervised representations},
  author={Choi, Hyeong-Seok and Lee, Juheon and Kim, Wansoo and Lee, Jie and Heo, Hoon and Lee, Kyogu},
  journal={Proc. NeurIPS},
  volume={34},
  pages={16251--16265},
  year={2021}
}

@inproceedings{siuzdak2022wavthruvec,
  author={Hubert Siuzdak and Piotr Dura and Pol {van Rijn} and Nori Jacoby},
  title={{WavThruVec: Latent speech representation as intermediate features for neural speech synthesis}},
  year=2022,
  booktitle={Proc. Interspeech 2022},
  pages={833--837},
}

@inproceedings{saeki2022utmos,
  author={Takaaki Saeki and Detai Xin and Wataru Nakata and Tomoki Koriyama and Shinnosuke Takamichi and Hiroshi Saruwatari},
  title={{UTMOS: UTokyo-SaruLab System for VoiceMOS Challenge 2022}},
  year=2022,
  booktitle={Proc. Interspeech 2022},
  pages={4521--4525},
}

@InProceedings{radford2022robust,
  title = 	 {Robust Speech Recognition via Large-Scale Weak Supervision},
  author =       {Radford, Alec and Kim, Jong Wook and Xu, Tao and Brockman, Greg and Mcleavey, Christine and Sutskever, Ilya},
  booktitle = 	 {Proc. ICML},
  pages = 	 {28492--28518},
  year = 	 {2023},
  volume = 	 {202},
  month = 	 {23--29 Jul},
  publisher =    {PMLR},
}

@article{chen2022wavlm,
  title={{WavLM}: Large-scale self-supervised pre-training for full stack speech processing},
  author={Chen, Sanyuan and Wang, Chengyi and Chen, Zhengyang and Wu, Yu and Liu, Shujie and Chen, Zhuo and Li, Jinyu and Kanda, Naoyuki and Yoshioka, Takuya and Xiao, Xiong and others},
  journal={IEEE Journal of Selected Topics in Signal Processing},
  volume={16},
  number={6},
  pages={1505--1518},
  year={2022},
  publisher={IEEE}
}

@article{sonobe2017jsut,
  title={{JSUT} corpus: free large-scale Japanese speech corpus for end-to-end speech synthesis},
  author={Sonobe, Ryosuke and Takamichi, Shinnosuke and Saruwatari, Hiroshi},
  journal={arXiv preprint arXiv:1711.00354},
  year={2017}
}

@inproceedings{lee2022bigvgan,
  title={{BigVGAN}: A universal neural vocoder with large-scale training},
  author={Lee, Sang-gil and Ping, Wei and Ginsburg, Boris and Catanzaro, Bryan and Yoon, Sungroh},
    booktitle={Proc. ICLR},
    year={2023},
}

@article{harvest,
title = "{{Harvest}: A high-performance fundamental frequency estimator from speech signals}",
author = "Masanori Morise",
year = "2017",
pages = "2321--2325",
journal = "Proceedings of the Annual Conference of the International Speech Communication Association, INTERSPEECH",
}

@article{morise2016d4c,
  title={{D4C}, a band-aperiodicity estimator for high-quality speech synthesis},
  author={Morise, Masanori},
  journal={Speech Communication},
  volume={84},
  pages={57--65},
  year={2016},
  publisher={Elsevier}
}
}

\end{document}